\documentclass[12pt,preprint]{aastex}
\begin{document}
\title{On the Nonspherical Nature of Halo Formation}
\author{Tzihong Chiueh}
\affil{Department of Physics, National Taiwan University, 
Taipei, Taiwan}
\email{chiuehth@phys.ntu.edu.tw}
\and
\author{Jounghun Lee}
\affil{Institute of Astronomy and Astrophysics, Academia Sinica,
Taipei, Taiwan}
\email{taiji@asiaa.sinica.edu.tw}

\newcommand{\etal}{{\it et al.}}

\begin{abstract}

We present a new collapse condition to describe the formation of
dark halos via nonspherical gravitational clustering.  This
new nonspherical collapse condition is obtained by the logical 
generalization of the spherical model to the nonspherical one. 
By solving  a diffusion-like random matrix equation 
with the help of the Monte Carlo method,  we show that this 
nonspherical collapse condition yields the mass function derived 
by Sheth \& Tormen (1999) which has been shown to be in excellent 
agreement with the recent N-body results of high resolution.  
We expect that this nonspherical collapse condition might provide us 
a deeper insight into the structure formation, and suggest that 
it should be widely applied to various cosmological issues such 
as the galaxy merging history, the galaxy bias, and so forth. 

\end{abstract}


\section{INTRODUCTION}

The mass function $n(M,z)$ in cosmology is defined to give the comoving 
number density of dark halos with mass $M$ at redshift $z$.  
It provides a useful analytical tool to understand the formation 
and evolution of the large-scale structure in the universe. 
The excursion set approach to the mass function provides the most 
direct and solid way to count the number density of dark halos.   
Bond et al. (1991) applied for the first time the excursion set theory 
to the Gaussian random density field, and recovered the popular
Press-Schechter mass function (Press \& Schechter 1974, hereafter PS) 
with a correct normalization factor of $2$.  

Sheth, Mo, \& Tormen (1999, hereafter SMT) suggested an extension of 
the excursion set approach to a nonspherical dynamical model. 
Although the PS mass function works fairly well at the high-mass 
section, recent hight-resolution N-body simulations have yielded 
less intermediate-mass and more low-mass halos than the PS  
prediction \citep{lac-col94,tor98,gov-etal99,jen-etal00}. 
It has been suspected that this discrepancy of the numerical 
mass functions with the PS prediction must be due to the 
departure of the true dynamics from the idealistic spherical one 
\cite{gov-etal99}. The work of SMT was in fact so motivated, 
attempting to justify the empirical  mass function derived by 
Sheth \& Tormen (1999, hereafter ST).   
Indeed, current numerical results from high-resolution N-body simulations  
agree with the ST formula much better than the standard PS mass function 
\cite{jen-etal00}.    

Yet it is hard to claim that the ST mass function is 
anything beyond a  phenomenological fitting formula.  
Although SMT claimed that the excursion set approach associated 
with their nonspherical collapse condition does produce the ST mass 
function to excellent approximation, a more careful analysis of the 
nonspherical collapse condition from a different perspective ought to 
be encouraged. In particular,  the perspective of symmetry and 
smoothness of the collapse condition  in the $\Lambda$-space (see $\S 2$) 
should be seriously considered, and can serve as a guiding 
principle in constraining the correct collapse condition. 

In this paper, we investigate the idea of SMT in a   
more sophisticated manner to find a physically motivated ellipsoidal 
collapse condition which yields almost the same ST mass function.   
The new collapse condition demonstrates in a clear-cut manner 
how the initial nonspherical properties of proto-halos affect 
the gravitational process, yielding the halo abundance substantially 
deviating from the PS predictions based on the spherical model. 

\section{NONSPHERICAL COLLAPSE CONDITION}

We use the following five hypotheses to evaluate the mass function:  
1) All cold dark matter elements eventually  
collapse into gravitationally bound halos by self similar clustering.  
2) The collaspe process is so rapid that the violent 
relaxation may completely erase the internal structure of 
the bound region.  
3) The collapse condition can be expressed by the linearly extrapolated
parameters of proto-halos.   
4) The rms fluctuations $\sigma(M)$ of the linear density field smoothed
over a mass scale of $M$ at the moment of collapse determines 
the mass of the bound halo.     
5) The collapse occurs in an ellipsoidal way, for which the 
necessary and sufficient collapse condition is a function of 
the three eigenvalues, $\lambda_1,\lambda_2,\lambda_3$  of the initial 
deformation tensor (defined as the second derivative of the linear 
gravitational potential). 

Note that the first four hypotheses are borrowed from the standard PS 
mass function theory.  The discrepancy from the PS theory 
arises in the fifth hypothesis.  For the spherical collapse model 
on which the PS theory is based,  the evolution of an initial 
spherical   overdense region is governed by the self gravity alone, 
so the collapse condition for the given region to form dark halos 
depends solely on its local average overdensity $\delta$ 
($\delta \equiv \Delta\rho/{\bar \rho}$. 
${\bar \rho}$: the mean mass density). 
While for the nonspherical collapse model,   
not only the self gravity but also the tidal interaction 
with the surrounding matter acts on the given region.   
Therefore, the evolution of the initial density inhomogeneities can 
be no longer described by the local average density alone once 
the simple constraint of the spherical symmetry on the initial region   
is released.  It must be described by other parameters quantifying  
both the intrinsic self gravity and the extrinsic tidal coupling with 
the neighbor mass distribution.  
It is worth mentioning here that it is this tidal interaction which 
causes the rotational motion of dark halos. That is, the generation 
of the angular momentum of dark halos is a unique consequence of the 
nonspherical collapse. 

Given the third hypothesis,  one can expect that the linear parameters 
to determine the {\it sufficient} nonspherical collapse condition may be 
the three eigenvalues of the tidal shear tensor 
(i.e., the deformation tensor), suggesting  the fifth hypothesis. 
Below we also explain why all the three eigenvalues 
determine the {\it necessary} nonspherical collapse condition.  
Our fifth hypothesis is qualitatively consistent with the peak-patch 
theory for the ellipsoidal dynamics proposed by Bond \& Myers (1996). 
SMT followed the peak-patch prescriptions to determine 
their nonspherical collapse condition.  However, we have noted that 
the collapse condition (eq. [3] in SMT) obtained from 
the peak-patch picture has some {\it unphysical} drawbacks.   
In our model, instead of relying on the peak-patch 
prescriptions fully, we suggest an original idea of the {\it logical 
generalization} of the spherical collapse condition into the 
nonspherical one (see $\S 4$).   

The above set of five basic hypotheses leads us to view the whole  
halo-formation process as a diffusion-like process of random fields 
in the three dimensional functional space spanned by the three eigenvalues  
$\lambda_1, \lambda_2, \lambda_3$ of the random deformation tensor.  
Note that here we use an {\it unordered} set of eigenvalues rather 
than the ordered one.  Thus, all three eigenvalues are equivalent. 
Let us consider a large-scale smoothed initial region where the local 
eigenvalues of the deformation tensor are given by 
$\Lambda = (\lambda_1, \lambda_2, \lambda_3)$, 
and the rms linear density fluctuations has a small initial value 
of $\sigma (M_0)$ with the corresponding mass scale of $M_0$. 
As one zooms in to look into the details, the rms fluctuations increases 
and  the probability distribution of $\Lambda$ \cite{dor70} becomes broader, 
making the low-mass bound objects easier to collapse.    
When $\Lambda$ of a given rms fluctuations $\sigma (M)$ just satisfies 
the collapse condition, this region will collapse into an 
{\it isolated bound halo} with mass $M$, according to the fourth 
hypothesis.  The isolated bound halo refers to the halo which has 
just collapsed with no larger halo enclosing it.  

Since the change of $\Lambda$ as the smoothing scale decreases is random, 
one can look upon this change of the smoothing scale as special kind of 
random walk of a particle in the three dimensional $\Lambda$-space. 
A particle corresponds to a bound region, 
and its position in the $\Lambda$-space is the local eigenvalues 
$\Lambda = (\lambda_1, \lambda_2, \lambda_3)$ of the deformation 
tensor defined at the region. 
Each random step is assumed to be independent, which amounts to choosing 
a sharp k-space filter to smooth out the density field 
\citep{pea-hea90,bon-etal91,jem95}. 
This random walk process is restricted within some absorbing boundary 
which corresponds to the collapse condition. 
The number of random-walk steps before the particle first hits  
the absorbing boundary is directly proportional to $\sigma (M)$, 
a decreasing function of M.  Thus, those particles that first hit  
the absorbing boundary in a small number of steps correspond to the 
high-mass halos while the opposite cases correspond to the low-mass halos. 

The shape of the absorbing boundary is determined by the collapse condition 
which is in turn governed by the underlying dynamics. 
For the case of the top-hat spherical model, the collapse condition 
is given by $\delta = \delta_c$ where $\delta_c$ is the density threshold.  
The original top-hat spherical dynamics gives $\delta_c \approx 1.69$ 
for a flat universe \cite{pee93}.  
However, the more realistic treatment of spherical collapse given the 
rapid virialization due to the growth of small-scale inhomogeneities 
gives $\delta_c \approx 1.5$ \cite{sha-etal99}.  In our approach, 
we use this realistic lowered value of $\delta_c$. 

Since $\lambda_1 + \lambda_2 + \lambda_3 = \delta$, 
the boundary for the spherical model is an infinite flat plane 
(the PS plane)  described by the equation of 
$\lambda_1 + \lambda_2 + \lambda_3 = \delta_c$ 
in the $\Lambda$ space.  Note that the PS plane is 
smooth (i.e., continuous and differentiable everywhere) 
and also symmetric about the line of $\lambda_1 = \lambda_2 = \lambda_3$. 
The latter property yields a rotationally invariant boundary, 
that is, the boundary remains unchanged under the exchange of 
the three variables ($\lambda_i \leftrightarrow \lambda_j$).  
We take these two important properties of the PS plane as the general 
properties that a physical collapse condition must satisfy.  
Given that the spherical collapse is a special case of the ellipsoidal  
one satisfying $\lambda_1 = \lambda_2 = \lambda_3$,  we expect that 
a physically meaningful ellipsoidal collapse condition must possess 
these two properties just as the spherical one does. 
The rotational invariance implies the isotropic nature of the 
initial matter distribution, while the smoothness implies  
the absence of any inherent singularity in the gravitational 
collapse as a general physical process. 

Note also here that these two required properties of the absorbing 
boundary,  the {\it rotational invariance} and the {\it smoothness},  
explains why the sufficient and necessary nonspherical collapse 
condition must be expressed in terms of all the three eigenvalues.  
If the absorbing boundary is expressed as a function of only one or two 
eigenvalues among the three, then either the rotational invariance or the 
smoothness must break in the $\Lambda$-space. 

At any rate,  it has long been pointed out that the spherical collapse
condition  is far from being realistic, and the true gravitational process
must be  ellipsoidal \cite{kuh-etal96}. Consequently the collapse condition
cannot  be expressed simply just by the density alone.  
Unfortunately, there has been no simple ellipsoidal dynamical
model that can describe the nonlinear regime adequately well, whereas 
the simplest top-hat  spherical model can trace all stages of halo 
formation even into the highly nonlinear regime after the moment of 
turn-around. The complicated nature of the tidal coupling with the 
surrounding matter in the nonlinear regime makes it extremely difficult 
to construct an universal ellipsoidal model from the first principles.   

Nonetheless, some qualitative considerations  on the nature of 
ellipsoidal dynamics can give us a hint for the collapse condition. 
The spherical collapse model is in fact a special case of the 
ellipsoidal one, satisfying $\lambda_1 = \lambda_2 = \lambda_3$. 
If the gravitational collapse of a bound region were spherical, 
then the differences between the three eigenvalues of a bound region, 
$|\lambda_i -\lambda_j|$ would remain zero during the collapse process.  
It implies that the the nonzero values of $|\lambda_i -\lambda_j|$ 
should quantify the {\it nonspherical} aspects of the true collapse. 
Using the given probability density distributions 
of each $\lambda$'s (see Appendix A in Lee \& Shandarin 1998), one can 
easily show that $|\lambda_i -\lambda_j| \propto \sigma (M)$ 
on average.   Thus,  as $\sigma$ increases, the average 
$|\lambda_i -\lambda_j|$ of a bound region also increases  
in proportional to $\sigma$.  Or, as the mass scale decreases, 
the degree to which the collapse deviates from the spherical model 
(the {\it nonsphericality}) increases.  
This explains why the PS mass function based on the top-hat spherical 
model works fairly well at the high-mass section where the 
nonsphericality is small \cite{tor98},  while it fails at the  
low-mass section where the nonsphericality is high. 
  
This idea can be quantitatively embodied by the $\Lambda$-space 
diffusion-like process.   As mentioned above, 
the spherical collapse condition is represented by an absorbing 
boundary of  an infinite flat plane in the $\Lambda$-space,  
with its distance to the origin being $\delta_c$. 
One may expect that the shape of the boundary for the nonspherical 
collapse should also be a smooth curved surface \cite{she-etal99}
that coincides with the PS plane at its apex, where 
$\lambda_1 = \lambda_2 = \lambda_3$.     
The underlying logic is as follows:  The distance from the PS plane 
to the nonspherical collapse boundary must provide a measure of the 
nonsphericality. The nonsphericality of the system is small at the 
high-mass section as argued above. In fact Bernardeau (1994) has shown 
semi-analytically that the evolution of rare events 
(very high-mass halos) is quasi-spherical. The formation 
of a high-mass halo therefore corresponds to the particles   
that reach the boundary near the $\lambda_1 = \lambda_2 = \lambda_3$ axis 
in only a small number of random-walk steps, and it suggests that 
the nonspherical collapse boundary should be smoothly tangential 
to the PS plane at the apex of that axis.  

Distant from the apex, the nonspherical collapse boundary 
should increasingly deviate from the PS plane. This is because 
the distant part of the boundary can be reached only by those 
particles that have undergone a large number of random steps, and  
they correspond to the low-mass halos, for which the nonsphericality 
is dominant. As the nonspherical collapse boundary should also possess
the smoothness and rotational invariance as 
the PS boundary does,  the analytic equation for this rotationally 
symmetric smooth surface can be constructed as follows.   
On the curved boundary, the ratio of  
$\delta (\equiv \lambda_1 + \lambda_2 + \lambda_3$) to $\delta_c$ 
equals unity only at the apex, $\Lambda_T$. This ratio  
becomes slightly larger than unity near $\Lambda_T$, 
and increasingly exceeds unity at distant points from $\Lambda_T$. 
Thus, the most general equation of the nonspherical collapse boundary  
can be written as  $\delta/\delta_c = S(r)$, where 
\begin{equation}
r \equiv \frac{1}{3}[(\lambda_1 - \lambda_2)^2 + 
(\lambda_2 - \lambda_3)^2 + (\lambda_1 - \lambda_3)^2].
\end{equation}  
and $(1/r)dS(r)/dr >0$.   Here the form of $r$ does guarantee 
the smoothness and rotational invariance of $S(r)$.  It is worth 
noting that the variable $r$ is proportional to the angular momentum 
square of the bound region provided that the principal axes of the 
inertia and the deformation tensors of the region are not perfectly 
aligned with each other \citep{hea-pea88, cat-the96}.  
Thus, the above general equation for the nonspherical collapse 
accounts for the generation of the rotational motion of dark halos.  

Specifically,  we propose the following boundary equation
for the nonspherical collapse: 
\begin{equation} 
\frac{\delta}{\delta_c}
= S(r) =  \left( 1 + \frac{r^2}{\beta} \right)^{\beta},
\end{equation}
where $\beta$  is a positive constant to be determined by fitting to the 
ST mass function.  Here note that the height of our absorbing 
boundary $S(r)$ scales like $r^2$ rather than $r$. It guarantees the 
flatter bottom of the absorbing boundary,   making it closer to the 
PS boundary around the apex, $\Lambda_T$ ($r=0$). Since the random walks 
representing the nonspherical collapse tend to avoid 
the axes of symmetry where any pair of the eigenvalues are the same
\cite{dor70}, the walks quicly diffuse away from $\Lambda_{T}$,  
never hitting the bottom of the absorbing boundary at $\Lambda_{T}$.   
Therefore in order to reproduce a mass function quite similar to the  
PS one at the high-mass section,  one needs an absorbing 
boundary with a flat bottom. We realize this flat-bottom absorbing 
boundary by expressing $S(r)$ scaled as $r^2$ rather than $r$. 

\section{ALGORITHM}

In this section, we describe the numerical algorithm for the 
Monte-Carlo simulation of the random-walk process in 
$\Lambda$-space given the above absorbing boundary (eq. [2]). 
This algorithm basically represents our  nonspherical collapse 
of bound regions into dark halos out of the initial Gaussian 
density field, which is in fact closely related to,  but simpler than 
the ellipsoidal collapse model developed by Eisenstein \& Loeb (1995). 
  
To simulate the initial random deformation tensor where the rms 
density fluctuations is $\sigma (M_0) \equiv \sigma_0$, we first 
generate six independent Gaussian variables with the dispersion of 
$\sigma_0$,  say $y_1, y_2, \cdots, y_6$.  
The symmetric deformation tensor $(d_{ij})$ can be constructed by the 
linear transformation of $(y_i)$ such that 
\begin{eqnarray}
d_{11} &=& -\frac{1}{3}\left( y_1 + \frac{3}{\sqrt{15}}y_2 + 
\frac{1}{\sqrt{5}}y_3 \right), \nonumber \\
d_{22} &=& -\frac{1}{3}\left(y_1 - \frac{2}{\sqrt{5}}y_3\right), 
\nonumber \\  
d_{33} &=& -\frac{1}{3}\left( y_1 - \frac{3}{\sqrt{15}}y_2 + 
\frac{1}{\sqrt{5}}y_3 \right), \nonumber \\  
d_{12} &=& d_{21} = \frac{1}{\sqrt{15}}y_4, \hspace{0.5cm} 
d_{23} = d_{32} = \frac{1}{\sqrt{15}}y_5, \hspace{0.5cm} 
d_{31} = d_{13} = \frac{1}{\sqrt{15}}y_6.
\end{eqnarray}
One can show easily that this linear transformation does satisfy the 
correlations of the deformation tensor \cite{bar-etal86}.   

Using the similarity transformation, we diagonalize the deformation 
tensor to find the three eigenvalues.  We then check whether the 
set of the three eigenvalues crosses the boundary or not.  If not, 
we generate a new six dimensional Gaussian random vector with the 
same dispersion of $\sigma_0$, and add it to the previous random 
vector.  Here the new random vector is assumed to be uncorrelated 
with the previous random vector, which is consistent with the use 
of the sharp k-space filter.   

Using this accumulated random vector, we repeat the above process: 
linear transformation into the deformation tensor, similarity 
transformation into a diagonal matrix to get the eigenvalues, 
and finally checking the boundary crossing.  This process is repeated 
until the first crossing over the boundary occurs. 
The number of the repetition ($N_s$) is proportional to the square  
of $\sigma (M)$ at the moment of  the halo formation such that 
$\sigma^{2} = N_s  \sigma_0^{2}$.  After a particle crosses the 
boundary, we re-starts the whole process with a new particle.   

We have simulated a ensemble of $120,000$ particles, and calculated
the  distribution of the number of particles which first cross the 
boundary at a range of $[\sigma, \sigma + d\sigma]$.  The results 
are plotted in Figure 1.  This distribution is nothing but the 
differential volume fraction  $dF/d\sigma$, occupied by the halos with 
the corresponding mass of $M$,   directly proportional to the mass 
function by $dF/d\sigma = [n(M,z)/\bar{\rho}]d{\rm ln}M/d\sigma$.  

To show the robustness of our approach, we also simulated the PS 
differential volume fraction by the above Monte Carlo method with the 
flat boundary. The triangle dots represent the resulting PS differential 
volume fraction  while the dashed line is the analytic standard PS formula.   
The numerical and analytical PS mass function agree with each other 
perfectly, which guarantees the robustness of the diffusion approach to
the mass function as well as the accuracy of our numerical scheme. 

\begin{figure}
\plotone{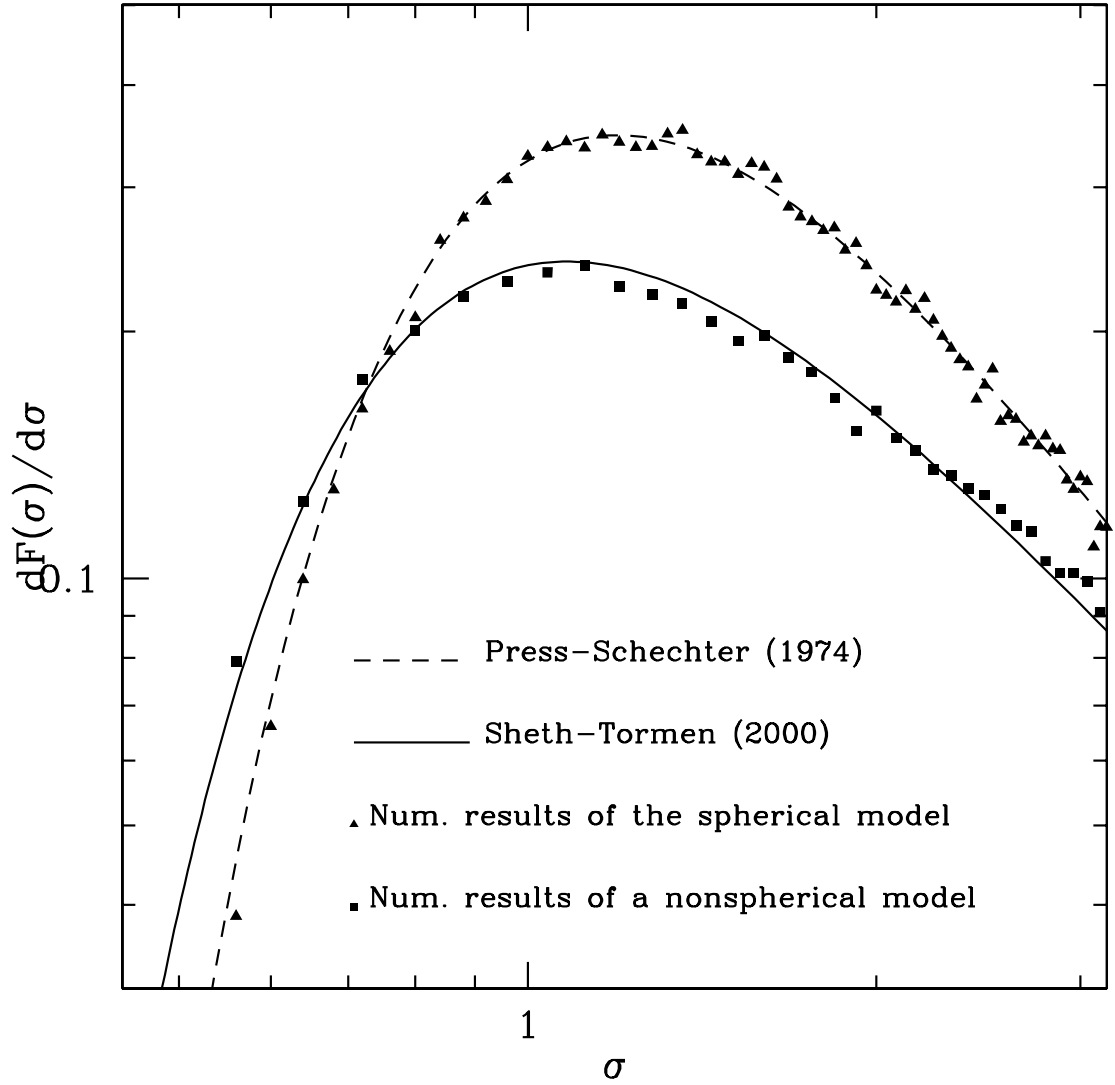}
\caption{The differential volume fraction from both the spherical and 
the nonspherical collapse conditions.  The fluctuation of the simulated 
mass functions are due to the numerical noise. \label{fig1}}
\end{figure}

The solid line is the ST formula which has been proved to fit 
the currently available N-body results of high resolution very well 
\cite{jen-etal00}, while the square dots represent our simulation 
results with the choice of the best-fit parameter of $\beta = 0.15$  
(and we used the realistic lowered value of $\delta_c = 1.5$ as 
mentioned in $\S 2$).   
Here we used the ST formula as our fitting standard to find this value 
of $\beta$. Of course, fine tuning of $\beta$ would be necessary if one 
is to use another fitting standard. 
As one can see, our result agrees strikingly well with the ST formula, 
suggesting that our collapse condition  can replace the PS spherical 
collapse condition in many interesting applications of the mass function. 

\section{DISCUSSIONS AND CONCLUSIONS} 

Motivated by the inspiring practical success of the ST mass function 
in recent N-body tests \cite{jen-etal00},  we have attempted here to 
provide a more sophisticated and robust way to determine the nonspherical 
collapse condition which produces the ST mass function to good 
approximation.  

In fact, various nonspherical approaches to the mass function were 
already attempted by several authors in the past decades 
\citep{mon95,aud-etal97,lee-sh98}.   
Strictly speaking, however, their approaches were not appropriate   
in the sense that they all used the original PS formalism 
which always yields ill-normalized mass functions.  Although the 
PS mass function can be properly normalized by multiplying a constant 
normalization factor of $2$ \cite{bon-etal91},  the mass function from 
a nonspherical model can no longer be corrected just by a constant 
normalization factor.  The normalization factor is scale-dependent in 
any nonspherical dynamical model,  due to the complicated pattern of the 
scale-dependent occurrence of the cloud-in-clouds.  
For a detailed description of the cloud-in-cloud problem,  
see Bond et al. (1991), Jedamzik (1995), and Lee \& Shandarin (1998). 
Due to this scale-dependent manner of the cloud-in-cloud occurrence 
in the nonspherical models, 
the shape of the mass function could be very different from the 
one obtained without considering the cloud-in-cloud occurrence correctly. 
For example, we have tested by our diffusion algorithm the condition 
that the bound objects form at the local maxima of the smallest 
eigenfield of the deformation tensor.  We found that the resulting 
mass function deviates considerably from the one originally given 
by Lee \& Shandarin (1998) who had used a constant normalization factor 
of $12.5$. 
 
In order to correctly solve the normalization problem, one should 
not rely on the original PS formalism. The only viable alternative 
for the evaluation of the mass function without extra efforts 
of concerning about the normalization is the excursion set approach, 
or equivalently the diffusion approach.  
SMT employed the excursion set approach to justify 
the ST mass function in terms of the nonspherical collapse condition. 
They related the nonspherical collapse condition with 
a moving barrier and approximated its shape with the help 
of the peak-patch prescriptions for the ellipsoidal dynamics 
\cite{bon-mye96}.  Their moving barrier (eq. [3] in SMT)  
is similar to our curved boundary in concept,  but differs significantly  
in practice since the SMT boundary is not a rotationally invariant 
smooth surface in the $\Lambda$-space and has kinks.  Given equation (3) 
in SMT, one can see that the kinks arise at $p = 0$ ($p$: the 
prolateness of the given ellipsoidal region, see Bardeen et al. 1986). 
As mentioned in SMT, however, $p = 0$ on average in a Gaussian random 
field. Thus, the kinks of the SMT boundary occur in so high probability 
regions that one may not ignore the presence of those kinks. 

Here we did not attempt to provide a better ellipsoidal dynamical 
model to describe the gravitational collapse process. Rather we 
retained the general framework of the peak-patch theory and try to improve 
the SMT boundary collapse condition into a more physically meaningful 
one by the logical generalization of the spherical collapse model. 
Two superior features of our collapse condition can be summarized 
as follows:  
First, it has a sound physical meaning, in that equation (2) 
is expressed in terms of only one single variable $r$ which is in 
fact directly proportional to the halo angular momentum 
square \citep{hea-pea88, cat-the96}.  Any physically meaningful nonspherical 
collapse condition should be expressed in terms of such quantities as 
obviously represent the nonshperical nature of halo formation.  
As mentioned in $\S 3$, the rotational motion of dark halos is a unique 
consequence of the nonspherical collapse. In this respect, our 
nonspherical collapse condition shows explicitly and quantitatively 
how the nonspherical gravitational collapse leads to the acquisition of 
the angular momentum of dark halos. Second it has the desirable 
analytical property, {\it rotational invariance and smoothness}.  
This feature of our collapse condition, in common with the PS 
collapse condition, will make it easy to extend various important 
cosmological issues, such as the halo merging, the light-to-mass bias, 
and so on, from the spherical dynamics to a nonspherical one.  

\acknowledgements 

We thank Z. Fan for helpful discussions and useful comments. 
We also thank our referee, R. Sheth, who helped us to improve the 
original manuscript.  
This work has been supported by the Taida-ASIAA CosPA Project.
T. Chiueh acknowledges the partial support from the National Science 
Council of Taiwan under the grant: NSC89-2112-M-002-065.

\newpage  
\end{document}